\documentstyle[pre,aps,float,epsf,multicol]{revtex}
\def\k{\kappa}

\def\l{\lambda}

\def\tsty{\textstyle}

\def\pr{\prime}

\def\br#1{\langle #1\rangle}

\def\b#1{{\bf #1}}

\def\cG{{\cal G}}
\def\cH{{\cal H}}

\def\vd#1#2{\b #1\kern -0.5mm\cdot\kern -0.5mm\b #2}
\def\mns#1#2{#1\kern -0.5mm - \kern -0.5mm #2}
\def\pls#1#2{#1\kern -0.5mm + \kern -0.5mm #2}
\begin{document}
\draft
\title{\bf Ginzburg Criterion for Coulombic Criticality}
\author{Michael E. Fisher and Benjamin P. Lee}
\address{Institute for Physical Science and Technology, University of
Maryland, College Park, Maryland 20742}
\date{April 12, 1996}
\maketitle
\begin{abstract}
To understand the range of close-to-classical critical behavior seen
in various electrolytes, generalized Debye-H\"uckel theories (that
yield density correlation functions) are applied to the restricted
primitive model of equisized hard spheres.  The results yield a
Landau-Ginzburg free-energy functional for which the Ginzburg
criterion can be explicitly evaluated.  The predicted scale of
crossover from classical to Ising character is found to be similar in
magnitude to that derived for simple fluids in comparable fashion.
The consequences in relation to experiments are discussed briefly.
\end{abstract}
\kern -0.1in
\begin{multicols}{2}

How can one understand the fact that some electrolyte solutions
display classical (or van der Waals) critical behavior down to
deviations from criticality of $|t|=|T-T_c|/T_c\sim 10^{-4}$ or less
\cite{MF}, while others exhibit purely Ising-type criticality
\cite{Ising} or, in some cases, {\it crossover\/} to Ising character 
at scales $t_\times\sim 10^{-1.5}$-$10^{-2.5}$ \cite{crossover,LeeFisher}?
The system triethylhexylammonium
triethyhexylboride (N$_{2226}$B$_{2226}$) in diphenyl ether \cite{MF} so far
reveals no hint of Ising character and is also the one that appears to
approximate most closely the simplest sensible theoretical model of an
ionic system, namely, the restricted primitive model (RPM), consisting
of $N=N_++N_-$ hard spheres of diameter $a$ with $N_+$ carrying
charges $+q$ and $N_-$($=N_+$) charges $-q$, in a medium of dielectric
constant $D$.  Real solutions and molten salts deviate from the RPM in
{\it many\/} ways: soft cores, differently sized ions, non-additive
ionic diameters, long-range van der Waals interactions, short-range
attractions, ionic polarizability, specific ionic chemical bonding,
the molecular structure of the solvent,
etc. \cite{FisherRev,StellRev,GG}.  At least some of these features
{\it must\/} be responsible for the observed {\it variation\/} of
$t_\times$, from $t_\times\simeq 1$ down to, perhaps, $t_\times\sim
10^{-5}$. Nevertheless, the most appealing scenario is that the RPM
itself, as an ``extremal model,'' displays purely classical critical
behavior {\it or\/}, failing that, has a very small value of
$t_\times$ that is increased by more realistic interactions.  (To
match the results for N$_{2226}$B$_{2226}$ one needs $t_\times\lesssim
10^{-4}$ \cite{FisherRev}.) Here we address this scenario.  Sadly, perhaps, 
our analysis, which encompasses all the leading physical effects
\cite{LeeFisher}, does {\it not\/} support this attractive picture!
Instead, it provides grounds for believing that $t_\times$ for the
RPM is of comparable magnitude to that for simple molecular fluids or
liquid mixtures that exhibit little if any non-Ising behavior.

To introduce our approach \cite{FL}, note first that all currently
available theories predicting ionic criticality (for dimensionalities
$d>2$ \cite{LLF}) yield classical behavior because, at heart, they are
of mean-field character \cite{FisherRev,StellRev,GG,FL,LLF}.  Recall also
that the standard approximate integral equations such as the HNC, YBG,
MSA, etc., yield {\it no\/} account of the critical region or else
fail to predict divergent critical-point density fluctuations
\cite{LeeFisher,FisherRev,FL}.  Second, direct simulations of the RPM
\cite{FisherRev,simulations} are far from being able to distinguish between
classical and Ising criticality.  In principle, a renormalization
group (RG) treatment of the fluctuations could reveal the true nature
of Coulombic criticality; but, in practice, that requires an
appropriate LGW effective Hamiltonian which has {\it not\/} been
available \cite{FisherRev}.  Furthermore, {\it quantitative\/} aspects
become important if, as seems likely [5,6(a)], a scale $t_\times$ is
present. 

However, as emphasized previously \cite{FL}, a sufficiently good
mean-field free energy can provide a foundation for an LGW Hamiltonian;
then one may estimate the domain of validity of classical critical
behavior by using the Ginzburg criterion \cite{Ginzburg} which,
indeed, is implied by RG theory \cite{MEF}.  If $m(\b r)=[\rho
(\b r)-\rho_c]/\rho_c$ is the order parameter (the overall ionic
density being $\rho\equiv N/V$) then, omitting the ordering field $h$,
the expected LGW form in $d$ spatial dimensions is
\begin{equation}\label{lgw}
{\cal H}/k_BT=a^{-d}\int d\b r[-\bar f(m)+{\tsty{1\over 2}}b_2^2
(\nabla m)^2+\ldots],
\end{equation}
with the spatially uniform reduced free energy density
\begin{equation}\label{fbar}
-\bar f={\tsty{1\over 2}}c_2tm^2-h_3tm^3+{\tsty{1\over 4}}u_4m^4
+O(tm^4,m^5). 
\end{equation}
The further gradient terms neglected in (\ref{lgw}) and the
corrections in (\ref{fbar}) are not needed for a leading order
description of criticality: {\it in principle\/}, however, large
or anomalous values of these terms could prove quantitatively significant.

To quantify the Ginzburg criterion we examine the  fluctuations of the
order parameter, normalized by the spontaneous order, $m_0$ ($t<0$), in
a $d$-sphere, $\Xi$, of radius set by the {\it correlation length\/}
$\xi(T)$, that is
\begin{equation}
\cG=\int_\Xi d\b r{\br{m(\b r)m(\b 0)}-\br m^2\over m_0^2}={\int_\Xi
d\b r G_{\rho\rho}(\b r)\over\rho_c^2 m_0^2|\Xi|_d},
\end{equation}
where $G_{\rho\rho}(\b r)=\br{\rho(\b r)\rho(\b 0)}-\br\rho^2$ is the
density-density correlation function, while $|\Xi|_{d=3}={\tsty
{4\over 3}\pi\xi^3}$.  Since $G_{\rho\rho}(\b r)$ decays fast on the
scale $\xi$, it is convenient to extend the integral to $\infty$ and
accept
\begin{equation}
\cG(T)=3\chi(T)/4\pi\rho_c m_0^2(T)\xi^3(T),\qquad(t<0)
\end{equation}
where $\chi=\int d\b r G_{\rho\rho}(\b r)/\rho$ is the reduced
susceptibility.  Treating (\ref{lgw}) and (\ref{fbar}) simply as a
mean-field free-energy functional, $F[\rho(\b r),T]$, yields the 
asymptotic relations
\begin{equation}\label{thermo}
\chi\approx{a^3\rho_c\over 2c_2|t|},\quad m_0^2\approx{c_2|t|\over u_4},
\quad\xi^2\approx{b_2^2\over 2c_2|t|},
\end{equation}
for $t\to 0-$ at $\rho=\rho_c$.  We may now set $\cG=1$ to obtain an
explicit estimate for the crossover scale $t_\times$, namely,
\begin{equation}\label{tGdef}
t_G=(9u_4^2/8\pi^2 c_2)(a/b_2)^6,\qquad\hbox{for }d=3,
\end{equation}
below which fluctuations dominate and the mean-field theory loses
validity:  Ising behavior should be exhibited for $|t|\ll t_G$.

Now, judging by concordance with the simulation estimates of $T_c$ and
$\rho_c$ [10,8(b)], the most successful available theory for the RPM
critical region \cite{FL} is based on the Debye-H\"uckel (DH) analysis
\cite{DH}, supplemented by (i) Bjerrum ($+,-$) ion pairing (Bj), (ii)
solvation of the dipolar pairs in the fluid of free ions (DI),
and (iii) hard-core
repulsions (HC) \cite{LeeFisher,comment}.  From the free energy
$F(\rho,T)$ of the DHBjDIHC theory (which is subject to minor
numerical variants [4,8(b)]) one can, by careful numerics
\cite{long}, extract, in addition to $\rho_c$ and $T_c$, $c_2$ and
$u_4$ in (\ref{fbar}); note that $h_3$ plays no role in (\ref{tGdef}).

But the Ginzburg analysis demands also the coefficient $b_2^2$ of the
gradient-squared term in (\ref{lgw}) that sets the amplitude of the
correlation length [see (\ref{thermo})].  Indeed, $b_2$ corresponds to the
{\it range\/} of the {\it effective\/} density-density attractions in
the RPM: these are embodied in the density correlation functions and,
more explicitly, in the wave-vector dependent susceptibility
\begin{equation}\label{chik}
\chi(\b k)={\chi(T)\over 1+\xi^2 k^2+\ldots}\approx{\rho_c a^d\over 
c_2t+b_2^2k^2+\ldots}.
\end{equation}
The last relation (for $t\to 0+$, $\rho=\rho_c$) follows by 
identifying, as before, the free
energy for a {\it nonuniform\/} system with ${\cal H}$
\cite{LeeFisher}.  In the past, DH theory has been
regarded as throwing light only on the {\it charge\/} correlation
function, $G_{qq}$, while remaining silent on $G_{\rho\rho}$.
However, in \cite{LeeFisher} we have shown how DH theory and its
necessary DHBjDIHC extensions \cite{FL} can be {\it generalized\/} in
a rather straightforward way to yield a classical functional of
$\rho(\b r)$ and thence an explicit mean-field expression for $\chi(\b k)$.
Furthermore, the unexpected divergence of $\xi(\rho,T)$ predicted when
$\rho\to 0$ turns out to be universal and {\it exact\/}!  By this route,
therefore, the critical coefficient $b_2$ is revealed and the
crossover scale $t_G$ can be explicitly calculated \cite{long}.  

Before describing our results for $t_G$, however, we note that
Leote de Carvalho and Evans \cite{LdCE} have recently demonstrated the
strategy set out above \cite{FL} by appealing to the generalized
mean-spherical approximation (GMSA).  This ingenious, OZ-based
\cite{comment} approximate integral equation \cite{gmsa} repairs the
simple MSA (for which the density fluctuations remain bounded) by
adding to the direct correlation functions a term with parameters
which are adjusted to satisfy various desirable sum rules; thence
$\xi$ diverges at criticality and $b_2$ can be estimated.
Unfortunately, however, the GMSA exhibits some serious defects: (a)
the correlation length $\xi(\rho,T)$ varies non-universally and quite 
incorrectly when
$\rho\to 0$ \cite{LeeFisher}, thence casting doubt on the plausibility
of the results for $\rho\simeq\rho_c$; (b) the predicted value of
$T_c$ is significantly too high \cite{FL,simulations}; (c) no account
is taken of Bjerrum pairing; and (d), apparently as a result of this,
the GMSA free energy violates Gillan's upper bound \cite{Gillan}
throughout the critical region (while DHBjDIHC theories satisfy it).
Consequently, although the GMSA values for $t_G$ \cite{LdCE}
provide an interesting benchmark, they are surely not adequate for the
purpose at hand.

For reference we start with  pure DH theory \cite{LeeFisher,FL} which
yields $c_2=1/64\pi$, $u_4=1/3072\pi$ \cite{fourpi} while $b_2^2/a^2=
(1+{20\over 3}\ln 2-6\ln{7\over 3})/64\pi\simeq(0.052)^2$
\cite{LeeFisher}.  Via (\ref{tGdef}) these yield $t_G\simeq 12.90$.
This number is large and certainly not suggestive of any significant
regime of classical behavior; but the derivation of (\ref{tGdef}) entailed
various essentially arbitrary numerical assignments.  For calibration,
therefore, it is essential to calculate $t_G$ by a comparable
procedure for a simple-fluid model that one can be confident exhibits
typical Ising behavior.

To that end we start with the functional generalization of the Mayer
expansion for a single-component fluid with a short-range pair
potential $u(r)$, namely,
\begin{eqnarray}
&F^{SR}&[\rho(\b r)]/k_BT=-\int d\b r\rho(\b r)[\ln\Lambda^3\rho(\b r)
-1]\nonumber\\
&-&{\tsty{1\over 2}}\int d\b rd\b r^\pr\rho(\b r)f\bigl(u(\b r-\b r^\pr)
\bigr)\rho(\b r^\pr)+O(\rho^3),
\end{eqnarray}
where $f(x)=\exp(-x/k_BT)-1$ and $\Lambda=h/\sqrt{2\pi mk_BT}$ [8(b)].  
This second-virial level
suffices to describe the attractions driving criticality, for which we
take a {\it square well\/} of  range $\l a$ and depth $\varepsilon$.  For the
repulsions we adopt {\it hard cores\/} of diameter $a$: to treat these we
follow our RPM approach \cite{FL} and approximate the $O(\rho^3)$
terms by a {\it local\/} expression of {\it free-volume\/} (FV) or
Carnahan-Starling (CS) form \cite{LeeFisher,FL}.  Note, however, that in
treating both this SqWHC model and the RPM, the CS expression
is not obviously preferable since the attractive interactions (direct
or effective) necessarily enter the true higher-order virial
coefficients and act to soften the hard core effects.

In the FV approximation the critical parameters are
\begin{equation}
\rho_c={\rho_{\rm max}\over 3},\qquad {\varepsilon\over k_BT_c}=
\ln\left(\l^3+57B^*/16\pi\over\l^3-1\right),
\end{equation}
with $B^*=1/a^3\rho_{\rm max}$.  For the range $\l=1.4$--$1.7$ that 
reasonably models real simple fluids \cite{McQuarrie}, $T_c$ depends 
strongly on
$\l$.  However, $k_BT_c=1.48\varepsilon$ should describe the corresponding
van der Waals/classical theory quite well \cite{FisherRev}: thus for
the assignments $B^*={\tsty{2\over 3}}\pi$ and ${\tsty{4\over 9}}
\sqrt 3$, which correspond to the exact hard-core second virial (2V)
coefficient and bcc close packing, respectively \cite{FL}, we choose
$\l\simeq 1.65$ and $1.43$.  The LGW parameters in (\ref{fbar}) are
found to be $u_4=3/16B^*$,
\begin{equation}
c_2=(57B^*+16\pi\l^3)\,\varepsilon/108k_BT_cB^{*2},
\end{equation}
\begin{equation}
b_2^2={2\pi a^2\over135B^{*2}}\left[{(\l^5-1)(1+57B^*/16\pi)\over\l^3-1} 
-1\right].
\end{equation}
Note that in the infinite range Kac-Baker limit, $\l\to\infty$ with
$T_c$ fixed, one has $b_2\to\infty$ and correctly finds $t_G\to 0$.

Using the CS hard-core form one can derive a quintic equation for the
critical density \cite{quintic} which yields $\rho_ca^3\simeq 0.249129$.  
Normalizing to $k_BT_c/\varepsilon=1.48$ as above leads to $\l\simeq 1.55$.

The third column of Table I lists the values of $t_G$ for the SqWHC
model found using the various hard-core approximations, with $\l$
chosen as indicated, and, for reference, with $\l=1.50$.  For
completeness, an RPA treatment \cite{LdCE} is included.  The
crucial LGW coefficients are also given.  The last column presents the
correlation length amplitude, $\xi_0^+$, defined via $\xi(\rho_c,T)
\approx\xi_0^+/t^{1/2}$ as $t\to 0+$ \cite{ampratio}.  These estimates all
lie quite close to $0.41a$.  On the other hand, the values of $t_G$
prove very sensitive to the approximations, ranging from $0.33$ to
$2.4$ (even discounting the HC/bcc value).  Since  $t_G\sim b_2^6$
[see (\ref{tGdef})] a strong dependence on $\l$ is not surprising; but
one might have hoped for better agreement among the approximate
methods.  Nonetheless, we may conclude that $t_G=10^{\pm0.5}$ will
characterize fluids that display only Ising behavior (or
`immediate' crossover). 

We may now assess the data given in Table II for the RPM.  In addition
to $t_G$ and $\xi_0^+/a$ (in the second and last columns) it is
instructive to examine the estimates for $T_c^*\equiv k_BT_cDa/q^2$
and $\rho_c^*\equiv\rho_ca^3$: these provide a measure of the merit of
the various approximations relative to the simulation data
\cite{simulations} which may be summarized by: $10^2T_c^*=5.2$--$5.6$,
$10^2\rho_c^*=2.3$--$3.5$ [8(b)].  Although the LGW coefficients here are
factors of 3--100 smaller than for the SqWHC model, the ratio
$\xi_0^+/a$ remains of order unity \cite{LeeFisher} and is again
fairly insensitive to the approximations: those yielding
($T_c^*$,$\rho_c^*$) in the simulation range suggest $\xi_0^+\simeq
0.80$.  For reference, the table also lists the critical values of
$Z\equiv p/\rho k_BT$ and the inverse Debye length $\k=(4\pi q^2\rho_1
/Dk_BT)^{1/2}$, where $\rho_1=\rho_++\rho_-=\rho-2\rho_2$ is the
density of {\it free-ions\/} while $\rho_2$ is that for the {\it ion
pairs\/} \cite{FisherRev,FL}.  The ratio $(\rho_2^*/\rho^*)_c$
measures the degree of pairing in the critical region: it is quite
significant \cite{FL,Gillan}.

The most striking feature of Table II, however, is the evidence that
$t_G$ for the RPM lies in the range $10^{0.3}$ to $10^{1.4}$ and so is
significantly {\it greater\/} than the value of $t_G$ for the
hard-core-square-well model!  From this perspective, the RPM should
{\it not\/} have an unduly small region of Ising-like character but
rather one of the same order, or even larger, than in simple fluids,
a conclusion certainly at  variance with the most natural
interpretation of the experimental evidence
\cite{MF,Ising,crossover,FisherRev}. 

\end{multicols}
\begin{table}
\caption{Ginzburg crossover scale, $t_G$, and critical parameters 
predicted for a hard-core square-well fluid (range $\l a$) [15].  See text
for hard-core (HC) approximations and Ref.\ [16] for RPA.}
\begin{tabular}{clllccccc}
HC & $\kern 6pt\l$ & $\kern 6pt t_G$ & $\kern -6pt k_BT_c/\varepsilon$ & 
$\rho_ca^3$ & $c_2$ & $u_4$
& $b_2/a$ & $\xi_0^+/a$ \\[0.1cm] \hline
bcc & $1.43_3$ & $0.09_7$ & $1.48$ & $0.433_0$ & $2.023_5$ & $0.243_6$
& $0.550_6$ & $0.38_7$\\
2V & $1.65_1$ & $1.62_1$ & $1.48$ & $0.159_2$ & $0.492_6$ & $0.089_5$
& $0.323_4$ & $0.46_1$\\
  & $1.50$ & $2.41_8$ & $1.13_1$ & $0.159_2$ & $0.539_5$ & $0.089_5$
& $0.298_0$ & $0.40_6$\\
CS  & $1.55_3$ & $0.28_4$ & $1.48$ & $0.249_1$ & $0.947_8$ & $0.113_1$
& $0.419_1$ & $0.43_0$\\
  & $1.50$ & $0.33_0$ & $1.33_4$ & $0.249_1$ & $0.979_2$ & $0.113_1$
& $0.406_6$ & $0.41_1$\\
RPA  & $1.50$ & $1.5_7$ & $1.26_7$ & $0.245_7$ & $0.673_5$ & $0.112_5$
& $0.333_2$ & $0.40_6$\\
\end{tabular}
\end{table}
\begin{table}
\caption{Ginzburg crossover scale, $t_G$, and critical parameters 
predicted for the RPM at various levels of approximations [4,15]: DH, pure
Debye-H\"uckel [4,13]; GMSA [16]; DHBj, with naive ion-pairing [5,8];
+DI with dipole-ionic fluid solvation (and $a_1=a$, $a_2=1.16198a$ [8]); 
hard-core treatments /bcc/2V/CS,
see text and [8]; DI$^\pr$, with a new charging process [23].}
\begin{tabular}{crllcclrrcl}
Approx. & $t_G\kern 6pt$ & $\kern -2pt 10^2T_c^*$ & $\kern -2pt 10^2\rho_c^*$ 
& $\k_ca$ & $10^2\rho_{2c}^*$
& $10Z_c$ & $10^2c_2$ & $10^3u_4$ & $b_2/a$ & $\xi_0^+/a$\\[0.1cm] 
\hline
DH  & $12.9_0$ & $6.25$ & $0.49_7$ & $1$ & $0$ & $0.90_4$ & $0.49_7$
& $0.10_4$ & $0.051_7$ & $0.73_3$\\
GMSA  & $1.0_8$ & $7.85_8$ & $1.44_8$ & $1.52_2$ & $0$ &  & $1.61_6$
& $0.33_9$ & $0.095_3$ & $0.75_0$\\
DHBj  & $12.9_0$ & $6.25$ & $4.51_7$ & $1$ & $2.01_0$ & $4.54_9$
& $41.01_4$ & $704.6$ & $0.469_4$ & $0.73_3$\\[0.2cm]
+DI  & $5.3_6$ & $5.74_0$ & $2.77_8$ & $1.12_3$ & $1.10_1$ & $2.23_6$
& $2.29_5$ & $1.53_0$ & $0.113_8$ & $0.75_1$\\
+DI/bcc  & $10.6_7$ & $5.54_2$ & $2.59_4$ & $1.02_9$ & $1.06_4$ & $2.48_4$
& $2.22_3$ & $2.24_9$ & $0.115_9$ & $0.77_7$\\
+DI/2V  & $23.2_5$ & $5.22_7$ & $2.44_3$ & $0.92_3$ & $1.04_4$ & $2.82_3$
& $2.17_8$ & $3.71_7$ & $0.120_8$ & $0.81_9$\\
+DI/CS  & $21.5_2$ & $5.24_9$ & $2.45_4$ & $0.93_1$ & $1.04_6$ & $2.79_8$
& $2.18_5$ & $3.54_8$ & $0.120_4$ & $0.81_5$\\[0.2cm]
+DI$^\pr$  & $1.2_3$ & $5.96_9$ & $2.38_1$ & $1.06_9$ & $0.91_9$ & $1.68_0$
& $1.77_0$ & $7.98_1$ & $0.122_3$ & $0.91_9$\\
+DI$^\pr$/bcc  & $2.2_5$ & $5.78_4$ & $2.14_5$ & $0.97_8$ & $0.85_2$ & $1.87_4$
& $1.65_3$ & $10.24_9$ & $0.121_5$ & $0.94_5$\\
+DI$^\pr$/2V  & $4.3_1$ & $5.50_6$ & $1.91_9$ & $0.87_7$ & $0.79_1$ & $2.13_7$
& $1.54_3$ & $14.08_0$ & $0.122_6$ & $0.98_7$\\
\end{tabular}
\end{table}
\begin{multicols}{2}

Although the DHBjDIHC theories account well for the leading physical
effects near criticality and all leading terms have been included in
$\cH_{LGW}$, it is {\it possible\/} that $t_G$, as  calculated here,
is a deceptive measure of the true RPM crossover scale, $t_\times$.
Perhaps the ion-dipole and dipole-dipole interactions, neglected as of 
order $\rho^3$,
play a special role \cite{GG}; this is being studied within a DH-style
approach \cite{charging}.  In principle, strong asymmetric terms, such as
$h_3$ and $u_5$ in (\ref{fbar}), could, under the full nonlinear RG flow,
invalidate the perturbative Ginzburg analysis.  Higher-order gradient
terms in $\cH_{LGW}$, especially if negative as arguments of
Nabutovskii {\it et al.\/} \cite{Nabutovskii} suggest, might, instead, bring
RPM criticality within the crossover domain of some multicritical
point \cite{FisherRev}.  The fact that the crossovers seen
experimentally are much sharper than standard \cite{FBB} (taking place
in a decade or less \cite{crossover,Anisimov}) supports this view.  To 
justify such a scenario,
however, seems to demand a more sophisticated and quantitative RG
analysis than normally feasible.

Conversely, if the RPM itself does exhibit {\it no significant crossover\/}
from classical behavior, {\it as our analysis indicates\/}, the anomalous
experimental results \cite{MF,crossover} must be ascribed to 
one or more of the features
lacking in the RPM that were listed initially \cite{FisherRev,StellRev,GG}.
Some of these, like the presence of short-range van der Waals or
solvent-mediated attractions, can and will be incorporated in our
formalism although, in truth, it is hard to see how they will
significantly alter the $t_G$ values.  Indeed, the experimental trends
seem, as mentioned, to indicate that $t_\times$ always {\it
increases\/} when the Coulombic forces have to compete with
solvophobic effects.  Unfortunately, one must also allow that
particular impurities might distort the data in unexpected ways: one
can imagine selective binding leading to `big' dipoles, or long ionic
`rods.'  The discovery, discussed in \cite{LeeFisher}, that the
dimensionless correlation length parameter $\xi_0^+\rho_c^{1/d}$ fits
our calculations rather well when the {\it Ising\/}-fitted amplitudes
are used may point in this direction.

Interactions with Professor R. Evans and Dr. R. J. F. Leote de Carvalho have 
been appreciated.  NSF Grant No. CHE 93-11729 supported our research.

\end{multicols}

\end{document}